# Simulating the deep decarbonisation of residential heating for limiting global warming to 1.5°C

**Florian Knobloch**[1,2], **Hector Pollitt**[2,3], **Unnada Chewpreecha**[3], **Vassilis Daioglou**[4,5], **Jean-Francois Mercure**[1,2]

[1] Radboud University Nijmegen, Department of Environmental Science, PO Box 9010, 6500 GL, Nijmegen, The Netherlands
[2] Cambridge Centre for Environment, Energy and Natural Resource Governance (C-EENRG), University of Cambridge, 19 Silver Street, Cambridge CB3 1EP, UK
[3] Cambridge Econometrics Ltd., Covent Garden, Cambridge CB1 2HT, UK
[4] PBL Netherlands Environmental Assessment Agency, PO Box 30314, 2500 GH, The Hague, The Netherlands
[5] Copernicus Institute of Sustainable Development, Utrecht University, Heidelberglaan 2, 3584 CS, The Netherlands

*Corresponding author:* Florian Knobloch, Email: f.knobloch@science.ru.nl, Tel. +31 (0)24 - 356 23 93

**Abstract:** Whole-economy scenarios for limiting global warming to 1.5°C suggest that direct carbon emissions in the buildings sector should decrease to almost zero by 2050, but leave unanswered the question how this could be achieved by real-world policies. We take a modelling-based approach for simulating which policy measures could induce an almost-complete decarbonisation of residential heating, the by far largest source of direct emissions in residential buildings. Under which assumptions is it possible, and how long would it take? Policy effectiveness highly depends on behavioural decision-making by households, especially in a context of deep decarbonisation and rapid transformation. We therefore use the non-equilibrium bottom-up model FTT:Heat to simulate policies for a transition towards low-carbon heating in a context of inertia and bounded rationality, focusing on the uptake of heating technologies. Results indicate that the near-zero decarbonisation is achievable by 2050, but requires substantial policy efforts. Policy mixes are projected to be more effective and robust for driving the market of efficient low-carbon technologies, compared to the reliance on a carbon tax as the only policy instrument. In combination with subsidies for renewables, near-complete decarbonisation could be achieved with a residential carbon tax of 50-200€/tCO$_2$. The policy-induced technology transition would increase average heating costs faced by households initially, but could also lead to cost reductions in most world regions in the medium term. Model projections illustrate the uncertainty that is attached to household behaviour for prematurely replacing heating systems.

**Conflicts of interest:** The authors declare that they have no conflict of interest.

**Acknowledgements:** FTT:Heat originates from a project commissioned by the European Commission, Directorate-General for Energy (Contract no. ENER/A4/2015-436/SER/S12.716128) (more details are available at the link http://ec.europa.eu/energy/en/data-analysis/energy-modelling). F.K. conducted the research and wrote the text, with support from H.P. and U.C. V.D. contributed the scenario inputs for future heat demand and contributed to the text. J.-F. M. coordinated the research and contributed to the text. The authors thank Lorena van Duuren and Pim Vercoulen for research assistance, Aileen Lam for data collection on Asia, Richard Lewney for coordinating the project from Cambridge Econometrics, Mark Huijbregts for feedback and valuable discussions, Joan Canton from the European Commission for guidance, other DG ENER staff for helpful feedback and additional guidance.

# 1 Introduction

Final energy demand in residential buildings was estimated to be 24PWh/y in 2010, causing direct on-site emissions of 2,18GtCO$_2$/y and indirect emissions from electricity use of 3,5GtCO$_2$/y (Lucon et al., 2014). At the same time, plenty of unexploited mitigation options exist in buildings, many at low (or even negative[1]) cost (Ürge-Vorsatz et al., 2012a). For limiting global warming to 2°C, all transformation pathways reviewed in IPCC-AR5 therefore suggest substantial reductions of the buildings sector's direct carbon emissions: around 50-75% until 2050, and up to 100% by 2100 (Clarke et al., 2014). Limiting global warming to 1.5°C implies a lower remaining budget for cumulative economy-wide emissions (730-880GtCO2 from 2015-2100, Millar et al., 2017), and therefore requires to reach net-zero carbon emissions as early as mid-century (Rogelj et al., 2015). Because remaining decarbonisation options compared to 2°C pathways are limited, scenarios for reaching the 1.5°C target rely much more strongly on rapid emission reductions in the buildings sector: around 85-95% by 2050, relative to their current level (Rogelj et al., 2015). While such pathways indicate what may be optimal from a social planning perspective, they strongly focus on the supply side of the energy system, leaving open the question how such a deep decarbonisation could be achieved by real-world demand-side policies, and which timescale would be realistic.

It is estimated that 56% of final energy in residential building is used for space and water heating, of which 55% is generated by fossil fuels, around 30% by biomass, and 15% by electricity and district heating systems (IEA, 2013a). Heating is thus by far the largest energy end-use in households, and responsible for most of residential on-site CO$_2$ emissions. Despite that, it only receives relatively limited policy attention compared to the electricity and transport sectors (IEA, 2014). While the theoretical emission reduction potential is well-known (Ürge-Vorsatz et al., 2012b), it remains unclear which real world policy instruments could reduce the sector's CO$_2$ emissions to near-zero sufficiently fast.

There is a consensus that the global demand for heating can be fulfilled much more energy-efficiently, thereby reducing fuel use and emissions without reducing comfort (Lucon et al., 2014). Heating loads can be reduced by an improved thermal insulation of houses, and the remaining heat demand can be serviced by renewable and energy-efficient technologies. Through their integrated application, building energy use can be reduced by up to 90% compared to conventional buildings (Urge-Vorsatz et al., 2013; Ürge-Vorsatz et al., 2012a). Given that 50% of the current building stock will still be in use by 2050 (75% in OECD countries) (IEA, 2013b), levels of building efficiency in the next decades strongly depend on building shell retrofits of existing houses (Ürge-Vorsatz et al., 2015b, 2012b).

Aside from space heating, over 40% of global heat demand is for water heating, which is less impacted by insulation (Connolly et al., 2014), but likely to rise with growing income in many world regions (Daioglou et al., 2012). Near-zero-emissions are thus unachievable as long as the remaining heat demand is not provided by renewable and efficient electricity-based technologies. Available alternatives to fossil fuel heating systems rely on the use of biomass (traditional or in modern boilers), electricity (e.g., electric resistance or immersion heating), ambient heat (heat pumps) or solar energy (solar thermal panels) (for an overview, see IEA, 2014). While the operation of solar and biomass systems can potentially be carbon-neutral (abstracting from life-cycle considerations[2]), heating with electricity can be a renewable technology once electricity generation is decarbonised, otherwise, electricity-related emissions must be accounted for. A much more efficient use of electricity can be achieved by heat pumps, which upgrade the ambient low-temperature energy of an air, water or ground source into higher-temperature heat for space and water heating, effectively achieving efficiencies of 200-400% (average ratio of heat output to electricity input) (for an overview, see IEA/ETSAP and IRENA, 2013). The cost-competitiveness of these capital intensive renewable technologies, with respect to incumbent fossil fuel technologies, strongly depends on a combination of local circumstances (such as local electricity and fuel prices, solar irradiation), generally not achieved in most countries.

---

[1] Cost is potentially negative when mitigation options do not only reduce emissions, but also the underlying energy use, which may lead to savings on energy expenses in excess of the necessary investment.

[2] Woody biomass can be considered as emission neutral assuming that plants grow to compensate the carbon emitted as they are combusted, so that the stock of forests (and carbon therein) does not decline as a result of biomass combustion. For discussions, see Zanchi et al. (2012) and Lamers and Junginger (2013).

The potential of modern renewable heating technologies remains largely unexploited: their combined use (excluding heat pumps) accounted for less than 5% of global heat generation in 2012 (for a market overview, see IEA, 2014). Some countries have introduced policy instruments and market and institutional conditions for incentivising their uptake, most commonly in the form of capital subsidies (e.g. in Austria and China), fossil carbon taxes (e.g. in Northern Europe), and use obligations (e.g. in Germany and Spain) (Connor et al., 2013; Ürge-Vorsatz et al., 2015a). Due to such policy support, an increased diffusion of renewables can be observed in some places (such as solar thermal heating in China): from 2010-14, their combined market share grew by 8% (International Energy Agency, 2017). While progress of the observed extent can improve the energy efficiency of heating incrementally, it is insufficient for achieving large absolute reductions in $CO_2$ emissions, which thus require more ambitious policy approaches (Jotzo et al., 2012; Mundaca et al., 2013; Mundaca and Markandya, 2016; Wilhite and Norgard, 2004).

Planning for a technological transition in the heating sector requires information on policy strategies that can generate the right level of incentive to achieve the required changes in consumer choices. In this paper, we focus on analysing the diffusion of renewable and efficient heating technologies in terms of various possible choices of realistic composite policy packages, in 59 regions of the world. Which policy mixes could induce a sufficiently fast transition towards renewable heating, under which conditions and behavioural assumptions, and how long would it take? In a first step, we use the IMAGE-REMG model (Daioglou et al., 2012; Isaac and van Vuuren, 2009) to project trends of residential heating demand until 2050. Our focus, however, is on the future technology portfolio, which depends on the choice of heating technologies by households.

We project household choices of heating technologies using the 'Future Technology Transformations' model FTT:Heat (Knobloch et al., 2017). It is a bottom-up simulation model of technology diffusion, aiming to project how the technology composition of residential heating systems in 59 world regions may develop until 2050 under the chosen assumptions on heat demand and choice behaviour. Based on projected preferences and decisions, the model simulates which technologies supply which share of the heating market, and estimates the resulting levels of fuel use and emissions. It is designed to simulate the potential impact of various sets of possible policy strategies.

The paper is structured as follows. In section 2, we review the literature, and in section 3 we present our model and data. Policy scenarios and results of the model simulations are presented and discussed in section 4. Section 5 concludes.

## 2  Background and literature

Residential heat generation is overwhelmingly small scale and distributed, taking place on site within homes. The uptake of new heating equipment depends on the individual decision-making by heterogeneous households, each with subjective preferences and perceptions, and only limited information, time and cognitive capacities to evaluate alternative options. At the system level, the sum of such decisions inevitably deviates from the least-cost optimum as it would be determined by models that assume a single, fully rational agent or social planner (Kirman, 1992), leading to the criticism that many bottom-up energy-models are based an unrealistic representations of real technology choice-making (for reviews, see Mundaca et al., 2010; Wilson and Dowlatabadi, 2007; Worrell et al., 2004). Avoiding costly policy-design failures requires an upfront simulation of policy effects, based on an analysis that better represents people's actual behaviour under bounded rationality and limited information, and accounting for nonlinear diffusion dynamics (Mercure et al., 2016; Rai and Henry, 2016). A behavioural modelling of decision-making is particularly relevant for policies aiming at a premature replacement of existing systems, which will likely be necessary for deep decarbonisation (Geels et al., 2017). For instance, in such decisions, households are found to apply very strict payback thresholds (Newell and Siikamki, 2015; Olsthoorn et al., 2017), reflected in high implicit discount rates as estimated from consumer choices (Hausman, 1979; Schleich et al., 2016; Sutherland, 1991; Train, 1985). In this section, we review the aspects relevant to building a simulation-based diffusion model that takes into account known behavioural features of decision-making.

## 2.1 Household decision making

Applied behavioural research shows that household decisions between different heating systems are driven by a diverse set of individual preferences and behavioural characteristics (for a review, see Kastner and Stern, 2015). Multiple studies find that main determinants of investments in heating systems are related to costs (Achtnicht and Madlener, 2014; Lillemo et al., 2013; Sopha et al., 2011). Other significant factors are the influence by social norms, and social comparisons with peers (Hecher et al., 2017; Sopha et al., 2010). Michelsen and Madlener (2012) find that preferences depend on socio-demographic (e.g. age, education), spatial (e.g. urban/rural) and home characteristics (e.g. home type, size). Wilson et al. (2015) point out that energy efficiency investments should not be seen in isolation, but as processes that emerge from social practices. Psychological and sociological research points to the behavioural relevance of norms, attitudes, values, motivation and social influence, all of which can impact technology choice (Abrahamse et al., 2005; Abrahamse and Steg, 2009; Clayton et al., 2015; Dietz et al., 2009; Steg, 2008; Stern, 1986; Wilson and Dowlatabadi, 2007).

The heterogeneity of household characteristics and perceptions may partly explain the 'energy efficiency gap' (Allcott and Greenstone, 2012; Jaffe and Stavins, 1994) between a hypothetical optimum as seen from an outside economics-engineering perspective and observed household decisions, which seem to undervalue future energy savings relative to upfront investment costs, resulting in a typically lower than expected uptake of energy-efficient technologies (reflected in high empirical estimates of implicit discount rates, as reviewed by Hausman, 1979; Train, 1985). Various explanations are discussed in the literature (for a review, see Gillingham and Palmer, 2014), which can be grouped into market barriers (e.g., hidden costs), market failures (e.g., asymmetric or imperfect information, split incentives) and systematic behavioural biases, as described by behavioural economics (e.g., loss aversion, suboptimal decision heuristics, status-quo bias, saliency, procrastination) (Allcott and Mullainathan, 2010; Bager and Mundaca, 2017; Frederick et al., 2002; Kahneman and Tversky, 1979; Lillemo, 2014; Shogren and Taylor, 2008; Simon, 1955).

The cumulative effect of household heterogeneity, imperfect information and social influence implies that the diffusion of new technologies does not happen instantaneously once they become economically attractive, but typically follows an S-shaped trajectory (Rogers, 2010): a slow initial growth is driven by a small group of early adopters, eventually followed by the large majority and, finally, laggards. The process is further amplified by learning-based gradual cost decreases once a technology grows in market share (Weiss et al., 2010), the nature and scale of transaction costs (Mundaca T et al., 2013), and industrial dynamics (e.g. capacity constraints, see Wilson, 2012). For energy technologies, these up-scaling dynamics have been extensively studied (Grübler et al., 1999; Wilson and Grubler, 2011), emphasising the significance of formative phases and path dependence. The resulting technology transitions usually take decades rather than years, being accompanied by changes in regulation, infrastructure, user practices and culture (Geels, 2002). Hence, policies aiming at transitions must be designed with an understanding of how new technologies slowly diffuse out of niches, and of households' diverse motivations to adopt and use them.

## 2.2 Modelling of the diffusion of heating systems

Various types of energy end-use models exists, which can be grouped into various categories, such as accounting-based, optimisation, agent-based or bottom-up (Mundaca et al., 2010; Worrell et al., 2004), each with different (or sometimes no) representations of decision-making (Wilson and Dowlatabadi, 2007). Many national or global level energy-economy models (such as Integrated Assessment Models) are of normative nature, used for analysing the cost-effectiveness (e.g., TIMES, MESSAGE) or cost-benefit ratio (e.g., FUND, PAGE, DICE) of technology pathways from a social planning perspective (Li et al., 2015).[3] While such an approach can be insightful for exploring cost-effective technology pathways, optimization models mostly neglect behavioural, social and industrial dynamics, effectively assuming that a single representative agent with perfect information and foresight is taking the decisions. The analysis of how (and if) a scenario can be achieved with which types of incentives requires simulation models that better capture some of the complex interactions between heterogeneous agents (Mercure et al., 2016), and some of the salient behavioural features involved in real decision-making, a need which is also highlighted in IPCC-AR5 (Kolstad et al., 2014).

For the heating sector, some attempts have been made to introduce a representation of household behaviour into optimization bottom-up models of technology uptake. Cayla and Maïzi (2015) extended the TIMES-Households model for France by including different household categories. Also for France, Giraudet et al. (2012) introduce consumer heterogeneity and intangible costs into a bottom-up sub-module of IMACLIM-R. Li (2017) presents an optimization model with heterogeneous decision-making which covers heating in the UK. However, most of these studies feature a relatively low resolution of technology types for heating, and are restricted to one (or mostly a few) countries. Furthermore, the behaviour represented typically comes in the form of frictions or externalities (e.g., using higher discount rates) introduced to prevent the models from reaching their otherwise normative optimal technological configurations.

As an alternative to optimization, several authors have developed agent-based models (ABMs) of heating system uptake, such as Sopha et al. (2011) for wood-pellet heating in Norway. While the approach allows the simulation of heterogeneous households, non-standard decision making and social influences, the calibration of ABMs requires a rich set of socio-economic data, usually generated by household surveys and interviews, which implies challenging requirements for their application on a larger scale (Rai and Henry, 2016). A more qualitative approach of analysis is taken in the 'Energy Efficiency in Buildings' model, which provides a replicable methodology for local market analysis (WBCSD, 2016).

A modelling approach of intermediate complexity is the use of discrete choice models with multinomial logit structures for representing household diversity, such as the CIMS model for energy demand in Canada (Rivers and Jaccard, 2005), the Invert/EE-Lab model for building-related energy demand in selected European countries (Kranzl et al., 2013; Stadler et al., 2007), and the bottom-up models of household energy use by Van Ruijven et al. (2011) and Daioglou (Daioglou et al., 2012) (together forming the REMG model component of IMAGE, see section 3.1), differentiating for urban and rural households and including income distributions. Some of these models are technology-rich and coupled with detailed building-physics models of the housing stock (e.g., Invert), which increases their degree of realism from an engineering perspective. At the same time, the data requirements are immense, so that models either tend to focus on a limited set of countries (CIMS and Invert), or only have limited technology representation (IMAGE-REMG).

---

[3] Under the incorrect assumption that the sum of individual cost-optimisers corresponds to a cost-optimising social planner, optimisation is often seen as descriptive, otherwise optimisation is always normative (Mercure, 2017).

Modelling the global scale is crucial in questions related to climate targets, as it is the only way to calculate global emissions required to estimate climate change. We acknowledge the difficulties to include (and parameterise) behavioural dynamics and diversity with only limited data available at the global scale, given that such behavioural features must vary by country. In FTT:Heat, we therefore choose a compromise: heterogeneous decision-making and social dynamics are represented in a stylised but tractable way. To the best of our knowledge, there exists so far no similar simulation model for the diffusion of heating systems at the global scale. This enables FTT:Heat to be made into a component of a new type of integrated assessment model, E3ME-FTT-GENIE (Mercure et al., 2018).

# 3   Methods and data
## 3.1   Simulating future heat demand with IMAGE-REMG

First, we use the IMAGE-REMG model to project future changes in heat demand ($UE_{tot}$), directly after the methodology described in Isaac and van Vuuren (2009) and Daioglou et al. (2012). Demand levels are projected for (i) a baseline scenario, (ii) a mitigation scenario consistent with limiting radiative forcing to 1.9W/m$^2$ (1.5°C target), involving increased efficiency of new buildings, and (iii) a variant of the mitigation scenario which additionally assumes increased retrofitting of existing houses.

IMAGE-REMG projects $UE_{tot}$ as the sum of demand for space and water heating. For water heating, future demand per person is modelled as a function of income, converging to a maximum saturation value which depends on heating degree days (HDD) (Daioglou et al., 2012). For space heating, demand is modelled as a function of population, floor space per person (*m2/cap*), heating degree days (*HDD*), and the useful energy heating intensity ($UE/m^2/HDD$) (Isaac and van Vuuren, 2009):

$$UE_{space} = Population \ x \ m^2/cap \ x \ HDD \ x \ UE/m^2/HDD \qquad (1)$$

Future changes in population, climate and income are exogenous drivers, based on the SSP (Shared Socioeconomic Pathway) 2 ('middle of the road') (see Riahi et al., 2017). The mitigation scenario (SSP2-1.9) projects reductions in heating intensity which would be consistent with achieving the 1.5°C target, which implies an improved thermal insulation of houses (e.g., by means of building standards and retrofits). All relevant data is publicly available via the IMAGE website (including future trends in m2/cap, HDD and $UE_{space}$) (PBL, 2018).

Floor space is an intermediate driver, calculated as a function of income and population density. HDD are derived from monthly mean temperatures, taking into account future levels of global warming. Heating intensity foremost depends on heating practices and levels of building insulation, with current values ranging from 50-200kJ$_{UE}$/m2/HDD. In both the SSP2 and SSP2-1.9 scenarios, the heating intensity of all houses is assumed to decrease towards an average of 90kJ$_{UE}$/m2/HDD by 2100 in all world regions (from 100-150 kJ$_{UE}$/m2/HDD by 2050), or remain at the current level if this is lower, meant to reflect the improvement of worst performing buildings with a convergence towards current average intensities. In the SSP2-1.9 scenario, it is additionally assumed that the adoption of more efficient insulation in newly built houses reduces the overall useful energy demand by up to 35%, compared to SSP2 (corresponding roughly to the "sub-optimal" scenario in the GEA, see Ürge-Vorsatz et al., 2012a). The retrofitting scenario is identical to SS2-1.9, but assumes that the heating intensity decreases towards 45kJ/m2/HDD by 2050, involving rapid insulation improvements of the existing building stock.[4]

## 3.2 Simulating technology uptake with FTT:Heat

The core of FTT:Heat is a simulation of technology diffusion, in which individual heating technologies (e.g., gas boilers, heat pumps) compete for market shares of the total heat demand. Importantly, the model does not minimise or maximise some objective function, such as system cost or intertemporal utility. Instead, it simulates the decision-making of households: under given behavioural assumptions and levels of heat demand, which heating technologies would they choose, and how fast can new technologies grow within the market?

Initial market shares of individual technologies $i$, $S_i(t)$, are calculated from historic data on heat generation by technology, $UE_i(t)$, as:

$$S_i(t) = \frac{UE_i(t)}{UE_{tot}(t)}, \qquad \sum_i S_i(t) = 1 \qquad (2)$$

At every time step $\Delta t$ of the simulation (set to 1/4 year), FTT:Heat models the change in future market shares based on three central elements (further described below):

1. Decision-making by diverse households
2. Dynamic growth of technologies
3. Learning by doing

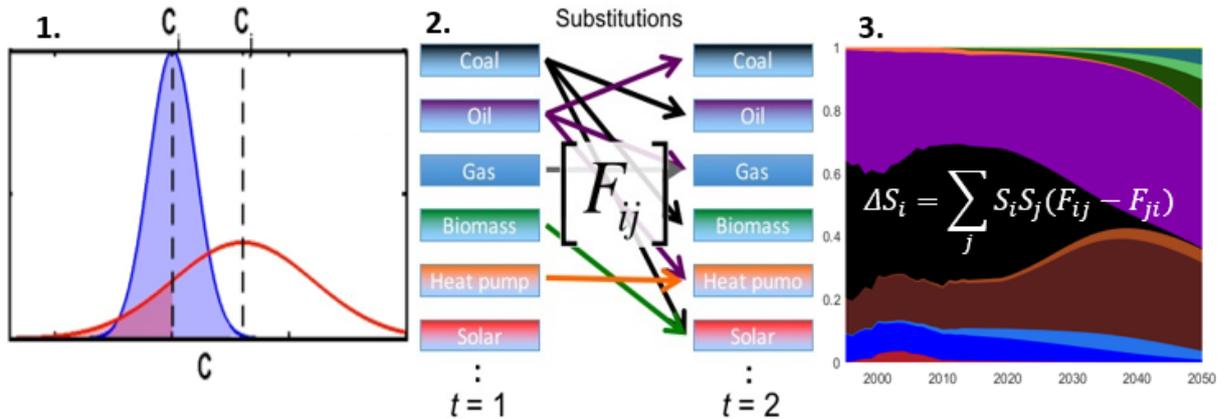

**Figure 1** An illustration of technology substitution over time in FTT:Heat. (1) Each period, in each region, some households decide between heating systems, based on distributed costs and intangibles. (2) All available technologies are compared with each other, and resulting household preferences stored in the matrix *F*. (3) The dynamic shares equation simulates the resulting net changes in each technology's market shares ($S_i$).

---

[4] The Passive House standard requires a maximum space heating energy demand of 15 kWh/m² (PHI, 2018). This roughly translates to a maximum useful heating intensity of 20 kJ/m²/HDD. As we model aggregate heating intensity, reducing global heating intensity from 50-200 to 45 kJ/m²/HDD over a 30-year period is very ambitious, and implies a large portion of households adopting passive house properties.

From the resulting shares, the model projects the new levels of useful energy demand per technology, $UE_i(t)$.

$UE_i$ can change when the technology composition ($S_i$) changes, and/or when the total demand for heating ($UE_{tot}$) changes. Final energy demand, fuel use and capacities are then estimated based on technology-specific conversion efficiencies and capacity factors (also depending on the climate). Negative changes in a technology's capacity correspond to decommissions that are not replaced. Finally, on-site $CO_2$ emissions are calculated from projected fuel use, based on the respective carbon content.

### 3.2.1 Decision-making by diverse households

In each period, a fraction of households decides between heating systems: either for replacing existing systems that come to the end of their lifetime (or became so expensive to operate that they are replaced ahead of that), or to satisfy new demand.

For systems coming to the end of their lifetime, FTT:Heat simulates a pairwise comparison of all available heating technologies by heterogeneous households, based on distributed costs and preferences. These are represented as the *generalised cost of heating*, $GCOH_i$:

$$GCOH_i = LCOH_i + \gamma_i \qquad (3)$$

$GCOH_i$ consists of two parts: an engineering-based levelised cost calculation ($LCOH_i$), and an empirical estimate of technology characteristics which are valued by households (based on observed technology uptake), $\gamma_i$. Levelised costs are calculated for all technologies, as:

$$LCOH_i = \sum_t \frac{\frac{IC_{i,t}}{CF_i} + \frac{MR_{i,t}}{CF_i} + \frac{FC_{i,t}}{CE_i}}{(1+r)^t} \bigg/ \sum_t \frac{1}{(1+r)^t} \qquad (4)$$

$IC_{i,t}$, $MR_{i,t}$ and $FC_{i,t}$ are upfront investment costs, maintenance-repair costs, and the fuel price respectively. $CF_i$ is the capacity factor (depending on a region's heating degree days), and $CE_i$ the technological conversion efficiency (output of useful heat, relative to input of fuel). $r$ is the discount rate. It is meant to describe how households value future relative to present costs, using a rate of 9%, based on Jaccard and Dennis (2006) (a sensitivity analysis is given in table 2)[5]. In addition, policies can be imposed, such as a carbon tax or a capital subsidy.

As a representation of household diversity, $IC_{i,t}$, $MR_{i,t}$ and $FC_{i,t}$ are all distributed around their mean values. Such diversity originates from different individual characteristics of the household, the technology or the dwelling. Accordingly, $LCOH_i$ is not treated as a unique value, but as a frequency distribution with a mean and a standard deviation, combined using standard error propagation:

$$\Delta LCOH_i = \sqrt{\frac{\Delta IC_i^2}{CF_i^2} + \frac{\Delta MR_i^2}{CF_i^2} + \frac{\Delta FC_i^2}{CE_i^2}} \qquad (5)$$

Many additional aspects may be valued by households which remain unspecified in the $LCOH_i$, such as the perceived inconvenience of a technology (e.g., for pellet heating, see Sopha et al., 2010), possible co-benefits (e.g., using a heat pump for cooling purposes), or existing policies, called the 'intangibles'. Their value is represented in the technology- and region-specific empirical parameter $\gamma_i$. It is derived using a calibration with historical diffusion data: we search for the set that makes the rate of diffusion continuous at the cross-over between historical and simulation periods, in each region (a description of the methodology is given in the appendix).

---

[5] ideally, we would have similar (and recent) empirical studies for all world regions. Unfortunately, those are not available. In their absence, some value needs to be chosen as a discount rate, so using results from the cited study on Canada is in our opinion an imperfect, but feasible solution.

For a population of heterogeneous households, the comparison of two technologies is then performed by comparing the frequency distributions of their generalised cost:

$$F_{ij}(\Delta C_{ij}) = \int_{-\infty}^{\infty} F_j(C) f_i(C - \Delta C_{ij}) dC, \quad \Delta C_{ij} = GCOH_i - GCOH_j \quad (6)$$

*F(C)* and *f(C)* are the cumulative cost distribution function and the cost distribution density, respectively. $F_{ij}$ denotes the fraction of households preferring technology *i* over technology *j*. It is calculated as the fraction of households for which the GCOH with technology *i* is less than with technology *j* (i.e., the model calculates a binary logit). For example, if $F_{ij} = 0.7$, 70% of households in this region would prefer technology *i* over *j*, while 30% have the reverse preference. The comparison is performed for all possible pairs of available technologies, resulting in a complete order of distributed household preferences, summarised in the matrix *F* (which is endogenously calculated for all regions, and updated in each simulation period). The diversity of choices implies a differentiation of the market, in which households take different decisions at different points in time for different reasons, as described by diffusion theory (Rogers, 2010).

### 3.2.2 Technology diffusion dynamics

Once preferences are estimated, the change in technology shares is simulated (Mercure 2015, 2012). First, we derive the substitution of market shares from heating technology *j* to *i* in period *Δt*, as:

$$\Delta S_{j \to i} = S_j F_{ij} \tau_j^{-1} S_i \Delta t \quad (7)$$

Substitutions from *j* to *i* depend on the fraction of households who would prefer technology *i* over *j* (the preference matrix $F_{ij}$, which is newly estimated in each period), the previous market shares of both technologies ($S_i$ and $S_j$), and the fraction of technology *j* which needs to be replaced (estimated as the inverse of its average technological lifetime, $\tau_j$)[6].

Since preferences are diverse, another fraction of households may have the reverse preference, and would choose technology *j* over *i*. We thus calculate the *net substitution* from technology *j* to technology *i*. Finally, the sum of all such pair-wise comparisons over all competing technologies *j* yields the *cumulative net change* in market shares of technology *i*:

$$\Delta S_i = \sum_j S_i S_j (F_{ij} \tau_j^{-1} - F_{ji} \tau_i^{-1}) \Delta t \quad (8)$$

---

[6] We assume homogenous age distributions because we do not have data over the age structure of heating systems worldwide. The assumption implies that if a fleet of heating systems of a particular technology has been widely adopted in a country in a very short time, it would mean that they would need replacement in a relatively restricted range of years, which would break our assumption. It does not imply, however, that new innovative systems could be replaced instantaneously; these would nevertheless wait until their end of life to be replaced (unless households would decide to replace prematurely). The constraint of not considering the exact age distribution is furthermore relaxed by the fact that systems have a probability of end of life that is distributed over time (not all systems of the same type end up with the exact same lifetime).

Formula (8) is the non-linear *dynamic shares equation*, conceptually similar to the modelling of imitation dynamics in evolutionary game theory (Hofbauer and Sigmund, 1998), originating in the description of evolutionary competition between different species. Each single flow from a technology *j* to an alternative technology *i* is determined by three interacting elements:

I. Preferences ($F_{ij}$): which fraction of households would prefer which technology, given that they were to buy a heating system within period $\Delta t$?
II. Replacement needs ($S_j \tau_j^{-1}$): how many heating systems of technology *j* need replacement in period $\Delta t$?
III. Dynamic constraints ($S_i$): given preferences and replacement needs, which fraction of substitutions can be realised?

Substitutions are dynamically constrained, as a stylised representation of (a) limited capacities to produce and install new technologies, and (b) limited information and behavioural decision-making on part of households. Psychological research reports that households typically don't optimise, but tend to stick to the status quo, are impacted by prevalent social norms, or follow the behaviour of others (Abrahamse and Steg, 2013; Frederiks et al., 2015). For heating systems, the main information sources are interpersonal sources and installers (Mahapatra and Gustavsson, 2008). For the diffusion of innovations, industrial and social dynamics are self-reinforcing, and both make it more likely that households choose dominant technologies. We represent these dynamics by constraining a technology's growth by its market share ($S_i$), based on Mercure (2017, 2015).

As a central implication, technology transitions in the model are subject to inertia, as technological trajectories cannot change direction rapidly, and resembles s-shapes diffusion curves. While undoubtedly remaining a stylised representation of underlying behavioural, social and industrial dynamics, it is an improvement in comparison to exogenous growth constraints in standard optimization models, since here the constraint is fully *endogenous*.

Substitutions can be further constrained to account for behavioural and technical plausibility/feasibility. We follow Kranzl et al. (2013) and assume that households would not switch back to technologies with a much lower comfort level, i.e. that households with modern heating systems (such as district heat, gas, electricity) would not go back to coal or traditional biomass.

Finally, new levels of heat generation per technology ($UE_i$) are obtained by multiplying their new shares by a region's total heat demand ($UE_{tot}$) (modelled by IMAGE-REMG).

### 3.2.3 Premature replacements

Eq (8) would suggest that households replace heating devices only at the end of their lifetime. However, in reality, households may consider to replace functioning heating systems ahead of that, based on economic considerations. For a household with perfect information and without risk-aversion, this would be beneficial once the marginal running costs of operating the current system exceed the full levelised costs of buying and operating an alternative technology. In practice, it is known that households apply much stricter criteria, and only consider a premature replacement if the potential savings exceed the initial investment in a limited period of time - the so-called *payback threshold* (Gillingham and Palmer, 2014).

Thus, in the model, premature replacements are only considered as sufficiently attractive if (and only if) the savings (due to reduced operating costs, $MC_i$) exceed the investment costs of another technology ($IC_j$, inclusive of an eventual subsidy) within the considered payback time (*b*), so if:

$$MC_i > MC_j + IC_j/b \qquad (9)$$

Given costs, decisions strongly depend on the assumed value of *b*. In choice experiments Olsthoorn et al. (2017) find that the mean payback threshold for a premature replacement of space heating systems is as low as 3±1 years, while Newell and Siikamäki (2015) report a mean threshold of 3-5 years. Assuming that this applies to all regions, we use 3±1 years. A sensitivity for *b* is given in section 4.4.

### 3.2.4 Learning by doing

We endogenously model cost reductions in upfront investment costs over time, $IC_i(t)$, which occur due to the accumulation of knowledge and experience ('learning by doing'), leading to the empirically well-described phenomenon of learning curves. Endogenous cost reductions increases path dependence in technology diffusion (Arthur, 1989), leading to increasing returns to scale for growing technologies.

In each period, updated investment costs are calculated as a function of the increase in cumulative global capacity of a technology, based on technology-specific learning rates: they describe the relative cost reduction that is expected for every additional doubling of the global capacity. We use learning rates from a review of empirical evidence by Weiss et al. (2010) (given in the appendix).

### 3.2.5 Economic feedbacks

FTT:Heat is hard-linked to the macroeconometric global simulation model E3ME (Cambridge Econometrics, 2014), most importantly via variables for fuel use and household expenditures. Policies which are primarily targeted at the heating sector can lead to changes in households' demand for different fuels, or to higher expenses for heating systems. For each year of the simulation, E3ME projects the wider macroeconomic impacts of such effects, and allows to analyse the implications for other economic sectors. We focus on the impact on electricity generation, and resulting indirect emissions in the power sector (which is modelled with FTT:Power, see Mercure et al., 2014).

## 3.3 Data

### 3.3.1 Energy demand

Only limited data is available on the specific demand for residential heating, the related fuel consumption and technology composition (IEA, 2014; Lucon et al., 2014). We combine data on fuel use and technology diffusion from various sources, and provide the resulting database as SI.

Final energy demand for heating in EU countries is taken from the ODYSSEE database (Enerdata, 2017), which contains annual data for space and water heating in all 28 member states. For non-EU regions, the IEA energy statistics report final residential energy demand by fuel type, but do not differentiate by end use application (IEA, 2017). The share of heating in household energy demand strongly depends on climatic conditions and levels of building insulation, ranging from 10% India to 87% in Russia, with water heating being of dominant importance in warmer climates (IEA, 2013a). We calculate the shares of heating for ASEAN, Brazil, China, India, Mexico, Russia, South Africa and the USA based on estimates in IEA (2013a) and country-specific sources (Daioglou et al., 2012; Eom et al., 2012; Wang and Jiang, 2017). For Sub-Saharan Africa, we adjust the fuel use data to bottom-up estimates of heat demand by the IMAGE-REMG model. For remaining world regions, the heating share is estimated based on heating degree days, assuming a comparable heating intensity as in world regions with a similar climate.

Residential heat generation by solar thermal installations for most world regions is available in the IEA energy statistics, which we amended by data from the IEA Solar Heating Programme (Mauthner et al., 2016). No standardised global data exists on heat generation by heat pumps. For most EU countries and Norway, time series were available from the European Heat Pump Association (2016), which was amended by data from EurObserv'ER (2017). For other world regions, the heat generation by ground-source heat pumps is taken from Lund and Boyd (2016). Data on the use of air-source heat pumps is taken from country-specific sources where available (China Heat Pump Committee of China Energy Conservation Association, 2015 for China; EIA, 2017 for USA; Japan Refrigeration and Air Conditioning Industry Association (JRAIA), 2017 for Japan; Kegel et al., 2014 for Canada; Lapsa et al., 2017 for the USA).

In the model, data on final energy demand ($E_i$) is transformed into useful energy demand ($UE_i$) according to technology-specific conversion efficiencies ($CE_i$) (see Appendix-Table ), so that:

$$UE_i(t) = E_i(t) * CE_i \qquad (10)$$

The resulting top-down estimate of heat demand has to be seen as a simplification, owing to the absence of reliable information on residential energy end-use in most world regions, and is subject to the uncertainty of underlying data and assumed efficiencies.

We estimate that in 2014, global final energy demand for residential space and water heating was around 12PWh, useful energy demand 9,6PWh, direct onsite $CO_2$ emissions 1,5GtCO$_2$/y (around 70% of reported direct residential building sector emissions in 2010, Lucon et al., 2014), and indirect emissions in the electricity sector 0,5GtCO$_2$/y.

### 3.3.2 Technology data

Cost and performance data for the 13 different kinds of heating technologies is summarised in Appendix-Table 2. Country-specific Investment costs (incl. of installation) for EU countries are taken from Fleiter et al. (2016) and Connolly (2014). For other world regions, we estimated the relative differences in investment costs from levels of real available household income. A standard deviation equivalent to 1/3 of the mean cost is assumed for all technologies (Danish Energy Agency, 2016, based on cost ranges reported by 2013; NREL, 2016). Residential fuel prices are taken from the IEA (2016a), with an assumed standard deviation of 15% (30% for biomass, based on NREL, 2016). More details and the data are given in the appendix.

### 3.4 Scenario definition

We created ten model scenarios (labeled **a-j**) aiming at a decarbonisation of residential heating until 2050, all of which use a different set of policies, implemented from 2020 onwards. Scenario **a** is our baseline projection, without improved insulation of the building stock (SSP2). It includes a continuation of current policies, the effect of which is implicitly included in the $\gamma_i$ parameters. Scenario **b** assumes increased levels of thermal insulation for new houses (SSP2-1.9). In scenario **c**, we simulate the additional effects of having more ambitious retrofitting of existing buildings, leading to further demand reductions. In scenarios **d-i**, we explore combinations of three policy instruments, in addition to improved insulation: a residential carbon tax, technology subsidies, and 'kick start' schemes for new technologies. These policies were chosen based on their successful previous implementation in at least some countries (for an overview, see Connor et al., 2013; IEA, 2014), as well as from advice from policy analysts at the European Commission. Scenario **j** explores complete electrification. We assume the political acceptability of the simulated policies for the purpose of our analysis. The gap between the modelled carbon tax levels and levels observed in reality reflects the political economy of carbon pricing mechanisms.

I. The (sectoral) *carbon tax* is specified as an absolute increase in the household price of fossil fuels, relative to their respective carbon content (we do not assume an inclusion of households into emissions trading).[7] We simulate carbon taxes of 50€/t$CO_2$ (scenario **d**) and 100€/t$CO_2$ (scenario **e**). From 2020-2050, the tax is assumed to linearly increase by +10% per year, relative to its respective starting value (reaching 200€/t$CO_2$ and 400€/t$CO_2$ in 2050). The 50€/t$CO_2$ tax is equivalent to fuel price increases of around +0.01€/kWh for gas, +0.013€/kWh for oil, and +0.018€/kWh for coal. In practice, such a tax may also take the form of subsidy removal in countries where household fuel use is currently subsidised (e.g., in form of reduced VAT rates, such as on domestic fuel use in the UK[8]).

II. *Technology subsidies* are defined as a relative reduction in a renewable heating technology's mean upfront investment cost. Eligible are solar thermal, heat pumps, and modern biomass. Two subsidy rates are simulated: -25% (scenario **f**) and -50% (scenario **g**). We assume that subsidy rates remain constant from 2020 until 2030, and are linearly phased out afterwards, reaching zero in 2050.

III. A *'kick start' policy* is not market-based, but represents a policy measure that introduces a new technology by means of a procurement scheme, use obligation in building codes, or other regulation. Such a policy is useful for driving initial markets in countries where a new technology's uptake is very limited or absent so far, thereby limiting the spread of relevant first-hand experiences between households, and preventing the build-up of necessary expertise and capacity in the building and heating industries. Precedents for such policies can be found in Spain and Germany, amongst others (Connor et al., 2013). In the model, we represent such a policy for limited time periods (5-10 years) by assuming that in each year, one percentage point of the dominant fossil fuel technology's market share is replaced by a mix of renewable alternatives.

Scenarios **d-g** focus on single policy instruments, while policy mixes involving two or more policies are simulated in scenarios **h-j**.

For all scenarios, we assume constant energy prices, for two reasons: first, future energy prices are highly uncertain, especially in a context of global deep decarbonisation. Effectively, this makes constant prices as likely as any other scenario. Second, it allows for a clearer identification of policy effects, which may otherwise be convoluted with the effects of a change in energy prices. A sensitivity analysis shows that our results do not substantially differ under increasing and decreasing future energy prices and other parameters (see table 2).

In all scenarios, solar thermal is limited to the demand of water heating in each country (as it is mainly used for water heating, and only as a supplementary source for space heating). District heating is only an option in those regions where it already exists (we do not assume the construction of new heat networks, which could well be an alternative to decentralised renewable heating in some regions).

---

[7] The specific carbon tax is only applied to the residential sector, and not assumed to be linked to other sectors, such as the power sector, which is subject to a separate set of policies.
[8] https://www.gov.uk/government/publications/vat-notice-70119-fuel-and-power/vat-notice-70119-fuel-and-power

# 4 Results
## 4.1 Heat demand

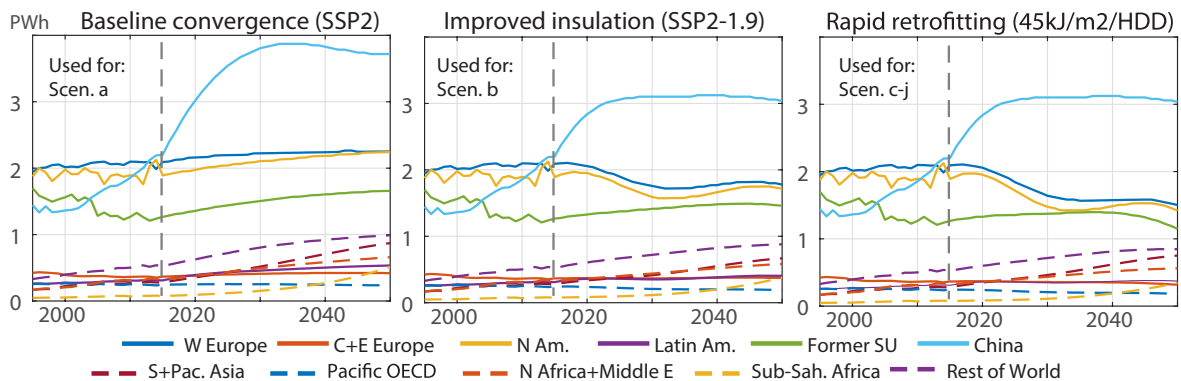

**Figure 2** Projections of future residential heat demand (space+water) by world region, for (a) a baseline scenario in which heating intensity converges to 90kJ/m2/HDD by 2100 (SSP2), (b) a scenario of improved building insulation of new houses in the context of climate policy (SSP2-1.9), and (c) a scenario which additionally assumes rapid thermal retrofitting of the existing building stock, so that the heating intensity converges to 45kJ/m2/HDD by 2050.

We first estimate future residential demand for space and water heating under different assumptions on future building efficiency and retrofitting, by applying projected demand trends from IMAGE-REMG to the historically estimated heat demand in 2014. Figure 2 depicts the resulting baseline demand trends by world region (SSP2, in which heating intensity globally converges to 90kJ/m2/HDD by 2100), a scenario of improved efficiency of newly built houses (SSP2-1.9), and a scenario which additionally assumes increased levels of retrofitting of the existing building stock (assuming a decrease to a heating intensity of 45kJ/m2/HDD by 2050). It becomes evident that changes in future global demand do strongly depend on China, which currently still shows very low levels of average heating intensity. Large demand increases are projected with continuously rising income, which may still be limited by improved building efficiency of newly built housing stock. Projected increases in warmer world regions mainly reflect growing demand for water heating, which empirically depends on income, and is unaffected by our assumptions on housing insulation. Estimated effects of retrofitting are largest in Western Europe and North America, where heat demand remains high, but is largely saturated.

## 4.2 Policies for decarbonisation

The main results for policy scenarios **a-j** are illustrated by Figure 3, which shows the projected global technology composition (left) and $CO_2$ emissions (right) until 2050. Indirect $CO_2$ emissions from electricity use are projected by FTT:Power, assuming a power sector decarbonisation scenario that is consistent with limiting global warming to 1.5C° (reaching zero emissions on the global level by 2040) (the emission trends for all regions are given in the SI). In addition, dashed lines show the total emission levels under baseline trends of power sector decarbonisation. Table 1 presents the cumulative changes in $CO_2$ emissions from 2020-2050. More details on the induced technology diffusion in five major world regions can be found in the appendix.

Values from 1995-2014 are estimates based on historical data, while the model simulation starts in 2015 (indicated by the dashed vertical line). 2014 values for total heat demand and emissions are represented as horizontal dashed lines. In plots for scenarios **b-j**, solid curves indicate the baseline demand trend (from scenario **a**). Percentage values refer to changes in demand and total annual $CO_2$ emissions by 2050, relative to 2014. Values in brackets refer to the respective changes in direct $CO_2$ emissions.

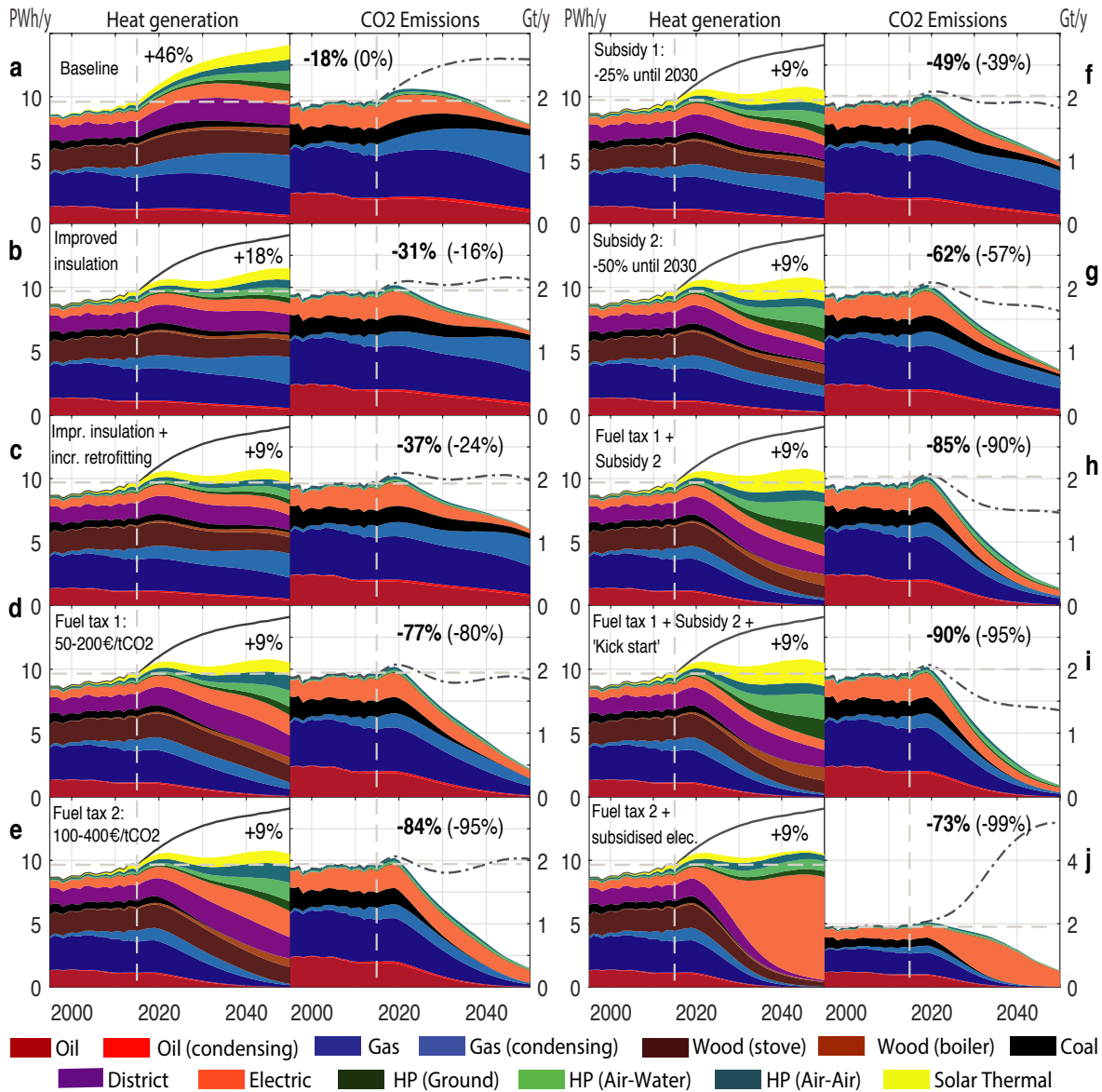

**Figure 3** Global technology composition and $CO_2$ emissions (direct on-site and indirect emissions from electricity use) in the residential heating sector, under current trends (a), improved building insulation (b), improved insulation and retrofitting (c), and seven policy scenarios aimed at technology uptake (d-j). Model simulations by FTT:Heat start in 2015 (indicated by vertical dashed lines). Horizontal dashed lines represent 2014 levels. Solid curves show the baseline demand trends from scenario a, and dashed curves the total emissions should the power sector not be decarbonised. Percentage values refer to the change by 2050, relative to 2014. Bold percentage values indicate the reduction in annual total $CO_2$ emissions (direct + indirect), the values in brackets show the corresponding reduction in direct on-site $CO_2$ emissions.

In our baseline projection under current policies and diffusion trends (scenario **a**), annual direct $CO_2$ would peak by 2030, before returning to their 2014 level by 2050. Given the projected increase in heat demand of +46%, this suggests a decrease in the direct emission intensity by around 30%. Importantly, these changes in our baseline do not result from any exogenous efficiency change, but from the endogenous continuation of current technology diffusion trends: the model projects a continuously increasing market share of heat pumps, solar thermal heating and modern biomass systems. Their combined share would grow from 9p.p. (percentage points) of heat production in 2014, to 30p.p. in 2050. An uptake of more efficient gas heating systems and a gradual replacement of coal and oil based systems leads to further emission reductions. Still, fossil fuel systems are projected to supply a more or less constant amount of heat, and keep a combined market share of 40p.p. by 2050, down from 50p.p. in 2014. Total emissions are projected to decrease by -18% by 2050, given that the power sector is decarbonised.

Table 1 Cumulative 2020-2050 $CO_2$ emissions, by residential heating and electricity generation for residential heating, in $GtCO_2$. The top row shows absolute values in the baseline scenario (a), the following rows the changes relative to baseline in scenarios with improved insulation and heating decarbonisation (b-j). Emissions by electricity generation are shown for a scenario with a power sector decarbonisation trajectory consistent with limiting global warming to 1.5°C (i), and a scenario without further decarbonisation of the power sector (ii).

|  | Scenario | Residential heating | Electricity generation for heating | | Heating + Electricity | |
|---|---|---|---|---|---|---|
|  |  |  | (i) Decarbonisation | (ii) Baseline | (i) Decarbonisation | (ii) Baseline |
| Baseline ($GtCO_2$) | a | 51Gt | 9Gt | 26Gt | 60Gt | 78Gt |
| Diff. to baseline ($\Delta GtCO_2$) | b | -8Gt (-15%) | -1Gt (-16%) | -5Gt (-19%) | -9Gt (-15%) | -13Gt (-16%) |
|  | c | -9Gt (-18%) | -1Gt (-16%) | +5Gt (+20%) | -11Gt (-18%) | -15Gt (-19%) |
|  | d | -25Gt (-48%) | +1Gt (+11%) | +6Gt (+21%) | -24Gt (-40%) | -19Gt (-25%) |
|  | e | -31Gt (-61%) | +3Gt (+39%) | +13Gt (+50%) | -28Gt (-47%) | -18Gt (-23%) |
|  | f | -13Gt (-26%) | -1Gt (-14%) | -4Gt (-16%) | -14Gt (-24%) | -17Gt (-22%) |
|  | g | -19Gt (-36%) | -1Gt (-11%) | -3Gt (-11%) | -20Gt (-23%) | -21Gt (-28%) |
|  | h | -30Gt (-59%) | +1Gt (+12%) | +4Gt (+14%) | -29Gt (-49%) | -27Gt (-35%) |
|  | i | -32Gt (-63%) | +1Gt (+11%) | +4Gt (+14%) | -31Gt (-52%) | -29Gt (-37%) |
|  | j | -36Gt (-71%) | +20Gt (+236%) | +77Gt (+292%) | -16Gt (-28%) | +40Gt (+52%) |

In scenario **b**, improved insulation of houses would reduce overall heat demand in 2050 by 20% (relative to baseline scenario **a**), indicating the untapped potential of increased building efficiency. In combination with continued diffusion of renewables, this could lead to a 31% reduction of total emission levels, although absolute heat demand would still be 18% larger than in 2014. Compared to baseline, increased thermal efficiency of new houses would reduce cumulative total $CO_2$ emissions by $9GtCO_2$.

In scenario **c**, we assume that all houses converge to an average heating intensity of 45kJ/m2/HDD until 2050, which implies that a large fraction of existing buildings would be retrofitted with improved insulation. As can be seen in Figure 2, this would foremost lead to additional demand reductions in North America, Europe and the Former Soviet Union (by 15-20% in 2050, relative to demand in SSP2-1.9). Total global heat demand by 2050 would be 9% lower than in 2014, leading to projected additional reductions of direct emission of a similar magnitude (compared to scenario **b**).

Other than for scenarios **b** and **c**, the resulting impacts of technology policies in scenarios **d-j** are not imposed by assumption (e.g., by assuming stricter building codes), but are subject to the simulated decision-making by households. The resulting technology transitions therefore depend on the behavioural characteristics of simulated decisions.

In scenario **d**, an (increasing) carbon tax of 50-200€/$tCO_2$ is introduced in 2020. Due to bounded rationality and limited information on part of households, the projected changes in technology uptake would only unfold very gradually, showing considerable inertia: households would still install new oil systems until 2035, and new gas systems even up to 2050. As households learn about the new technologies, and industry capacities expand, renewables would increasingly grow in market shares. Compared to scenario **c**, annual heat generation in 2050 by heat pumps would be 90% larger, solar thermal by 18%, and modern biomass by 50% (reaching a combined market share of 45p.p. by 2050). The resulting decrease in annual total emissions would be 77% in 2050

In scenario **e**, an (increasing) carbon tax of 100-400€/t$CO_2$ is projected to bring down direct emissions by 95%, and total emissions by 84%. Relative to the lower carbon price in scenario **d**, the 2050 market shares of heat pumps and modern biomass would increase by +13% and +11%, while shares of solar remain virtually unaffected. This indicates a decreasing marginal impact of carbon taxes on household decisions: although moderate tax levels can be sufficient for slowly steering household choices for regular replacements and new installations away from fossil fuel technologies, slow turnover rates limit the resulting pace of change in the technology composition. Furthermore, households tend to discount future fuel savings which can be achieved by adopting highly efficient, but capital-intensive modern renewables. Such technologies are disadvantaged by their higher upfront costs, which have a more than proportional impact on households' decision-making. Instead, the carbon price would induce a shift towards (less efficient) direct electric heating (which would grow by 30%, relative to scenario **d**), resulting in indirect emission increases.

Subsidy schemes are simulated in scenarios **f** and **g**. The projected decrease in annual total $CO_2$ emissions by 2050 is 49% for a 25% subsidy, and 62% for a 50% subsidy. Results suggest a clear shift of household choices towards more capital-intensive and efficient technologies, in comparison to the carbon tax: solar thermal and ground-source heat pumps show the largest increases in uptake (up to +100% and +200% relative to scenario **c**, respectively). Accordingly, electricity use for heating is around 50% lower than in the carbon tax scenarios, making the technology portfolio much more robust against potential indirect emission increases in the power sector. Still, results suggest that on their own, even large subsidies could not motivate many households to replace functioning systems prematurely. Even if 'scrapping' would look beneficial from an outside perspective, households tend to apply strict behavioural payback criteria for such decisions, which makes 'scrapping' unattractive from their subjective perspectives (see section 3.2.3).

In scenario **h**, the 50% subsidy on renewables is combined with the 50-200€/t$CO_2$ carbon tax. From a behavioural perspective, the policy mix impacts households' decision-making like 'carrot and stick': the tax enacts a steadily increasing economic pressure on households, strong enough so that a growing proportion of them may eventually want to reconsider their preference for fossil fuel technologies. In parallel, the subsidy has an over-proportional impact on costs as they are perceived by households, in particular in case of premature replacements (see Figure 5), thereby benefiting modern renewables. According to the model projections, the policy mix could thus induce a wave of premature replacements. As a result, annual total emissions by 2050 would be reduced by 85%.

While the policy mix of scenario **h** is projected to incentivise a decarbonisation pathway in most world regions, results suggest that the policies would be relatively ineffective in the Middle East and Russia, where two-third of remaining emissions would occur by 2050. In both regions, the diffusion of renewables is not just constrained by comparably low fossil fuel prices, but also by the practical absence of such technologies in the present technology mix, which implies limited local knowledge and experience. Scenario **i** therefore adds 'kick start' policies for driving the initial market for renewables in Russia and OPEC states, e.g. in form of procurement schemes (see section 3.4). Once households and local industry learn about the new technologies, a diffusion process is nucleated, and the financial incentives can have a much larger impact. Overall, the policy mix is projected to reduce total emissions by 90% (direct emissions by 95%), the largest total reduction in all scenarios.

Results of scenarios **h** and **i** imply that mixing policies enables to impose a lower carbon price and reduce costs to households (see section 4.3), in contrast to common model assumptions in which all policies are assumed to have a 'carbon price equivalent'. In a diffusion model, policy interacts, and the behavioural details of the policy mix (and how it impacts household decisions) matters. Compared to a carbon price on its own (scenario **e**), our results suggest that the transition to renewables could take place with more efficient (albeit more capital intensive) technologies: although leading to similar direct emission reductions, the carbon tax on its own would lead to much higher electricity use, leading to larger expenses on electricity and indirect emission increases in the power sector.

In all scenarios, an increased electrification of residential heating has direct implications for the power sector. Annual electricity demand for residential heating in 2014 was around 1PWh/y (or 4% of global electricity demand), causing indirect emissions of 0,5GtCO$_2$/y. In our baseline projection, it would increase to 2PWh/y in 2050. Without decarbonisation of the power sector, indirect emission could then reach 1,1GtCO$_2$/y. Net reductions in cumulative 2015-2050 CO$_2$ emissions thus strongly depend on a parallel decarbonisation of electricity generation (see table 1). Indirect emission increases cancel out 5-20% of direct emission savings when assuming a rapid power sector decarbonisation, but up to 55% when the sector continues on its current trajectory. Net savings are much more sensitive to induced power sector emissions in scenarios which only rely on a carbon tax, due to their relatively higher levels of electricity demand. Despite this, residential emission reductions in scenarios **c-i** would always exceed potential emission increases in the power sector.

Table 2 Sensitivity of cumulative 2015-2050 direct CO$_2$ emissions in residential heating, with regard to: (i) future energy price developments (relative to scenarios with constant future energy prices), assuming: (a) linear increase in all residential fuel prices by +1% per year (2018-2050), (b) linear decrease in all residential fuel prices by -1% per year (2018-2050); (II) learning rate for heating systems (50% lower than default assumption); (III) discount rate (50% higher than default assumption); (IV) 'intangible' component of generalised heating cost (50% lower than default estimates).

| | % deviation of cumulative 2015-2050 direct CO$_2$ emissions, relative to default parameters | | | | |
|---|---|---|---|---|---|
| | **Increasing fuel prices** (linear by +1%p.a.) | **Decreasing fuel prices** (linear by -1%p.a.) | **Learning rate** (-50%) | **Discount rate** (+50%) | **Intangibles** (-50%) |
| **a** | -4% | +5% | +3% | +3% | +7% |
| **b** | -3% | +4% | +2% | +2% | +7% |
| **c** | -3% | +4% | +2% | +2% | +7% |
| **d** | 0% | +1% | +4% | +2% | +10% |
| **e** | +2% | -1% | +4% | +1% | +12% |
| **f** | -4% | +6% | +3% | +2% | +8% |
| **g** | -3% | +7% | +5% | +1% | +10% |
| **h** | 0% | 0% | +5% | +1% | +10% |
| **i** | -1% | +1% | +6% | +1% | +9% |
| **j** | +3% | -2% | +2% | +1% | +25% |

In scenario **j**, we explore the extreme case of complete electrification. In addition to the 100-400€/tCO$_2$ carbon tax from scenario **e**, we assume that electricity use for heating is subsidised by 0,05€/kWh, and the purchase of all electricity-based systems by 30%. Because electric heaters are a readily available and well-known technology in most world regions, also being clean and convenient in use, our model suggests a relatively rapid uptake once costs are favourable: direct electric heating alone would gain a 75% market share by 2050, reducing on-site emissions by as much as 99%. However, resulting electricity demand would reach 9PWh/y, requiring 2.000-2.400GW of additional generation capacity (almost half of the currently installed global capacity). Indirect emissions would cancel out 80% of direct $CO_2$ reductions even under power sector decarbonisation, and could reach up to 6$GtCO_2$/y otherwise. This makes a direct electrification of heating an overall rather ineffective (or even counterproductive) mitigation strategy, particularly given the availability of much more efficient alternatives.

Table 2 presents the sensitivity of our scenario results with respect to key parameters (for the sensitivity with respect to 'scrapping' assumptions, see section 4.2). It becomes evident that results are relatively robust to changes in cost parameters: under alternative assumptions on fuel price trends, learning rates and discount rates, cumulative direct $CO_2$ emissions do not change by more than 7% in any scenario. The largest deviations can be seen for changes in 'intangible' cost components: if they are reduced by 50%, cumulative emissions are up to 12% higher (even 25% in case of the complete electrification scenario, because electric heating would become less attractive). This is partly due to the fact that large 'intangible' costs are attributed to coal, the reduction of which would lead to the uptake of emission-intensive coal heating systems.

### 4.3 Cost-effectiveness of policy mixes

From a public policy perspective, the decarbonisation of heating could not only be beneficial for climate change mitigation, but potentially also enable a more efficient provision of heat in monetary terms. The cumulative projected costs and savings from the induced technology transitions (in scenarios **d-j**) are presented in table 3, relative to scenario **c** (with improved insulation, but without technology policies). Figure 4 exemplarily shows the underlying changes over time per world region in case of four scenarios.

Table 3 Cumulative changes in global expenses on residential heating (billion Euro, bn€), for policy scenarios d-j (aiming at uptake of low-carbon technologies), relative to scenario c (improved insulation and retrofitting, without technology policies). Monetary values are shown in constant 2015-Euros (undiscounted). Changes in expenses on heating systems per t$CO_2$ refer to net reductions in $CO_2$ emissions (direct plus indirect from electricity use).

|  | Scenario | Expenses on heating systems | | Expenses on energy for heating | Total expenses on heating | Policy revenue (Tax minus subsidy) |
|---|---|---|---|---|---|---|
|  |  | bn€('15) | €('15)/t$CO_2$ | bn€('15) | bn€('15) | bn€('15) |
|  | c | 4.600 | --- | 16.030 | 20.630 | --- |
|  | d | +460(+10%) | 36 | -1.010(-6%) | -550(-3%) | 2.750 |
|  | e | +770(+17%) | 44 | -830(-5%) | -60(-0%) | 3.730 |
|  | f | +460(+10%) | 121 | -1.160(-7%) | -700(-3%) | -400 |
| Diff. to (c), in bn€('15) | g | +1.170(+25%) | 132 | -2.600(-16%) | -1.430(-7%) | -1.220 |
|  | h | +1.540(+33%) | 82 | -3.190(-20%) | -1.650(-8%) | 590 |
|  | i | +1.650(+36%) | 80 | -3.400(-21%) | -1.750(-8%) | 240 |
|  | j | +1.640(+36%) | 284 | +3.080(+19%) | +4.720(+23%) | -10.800 |

In all scenarios, cumulative expenses on heating systems are projected to increase between 10-36%. Subsidies would lead to relatively larger increases (compared to carbon taxes) in terms of €/tCO2 net reduction, as they incentivise the purchase of more capital-intensive technologies. At the same time, the new technologies are much more energy-efficient, thus enabling substantial reductions in energy expenses for heating, which range between 5-21%. As for expenses on heating systems, policy mixes which involve subsidies are also projected to result in larger energy savings, as they lead to the adoption of more energy-efficient technology portfolios.

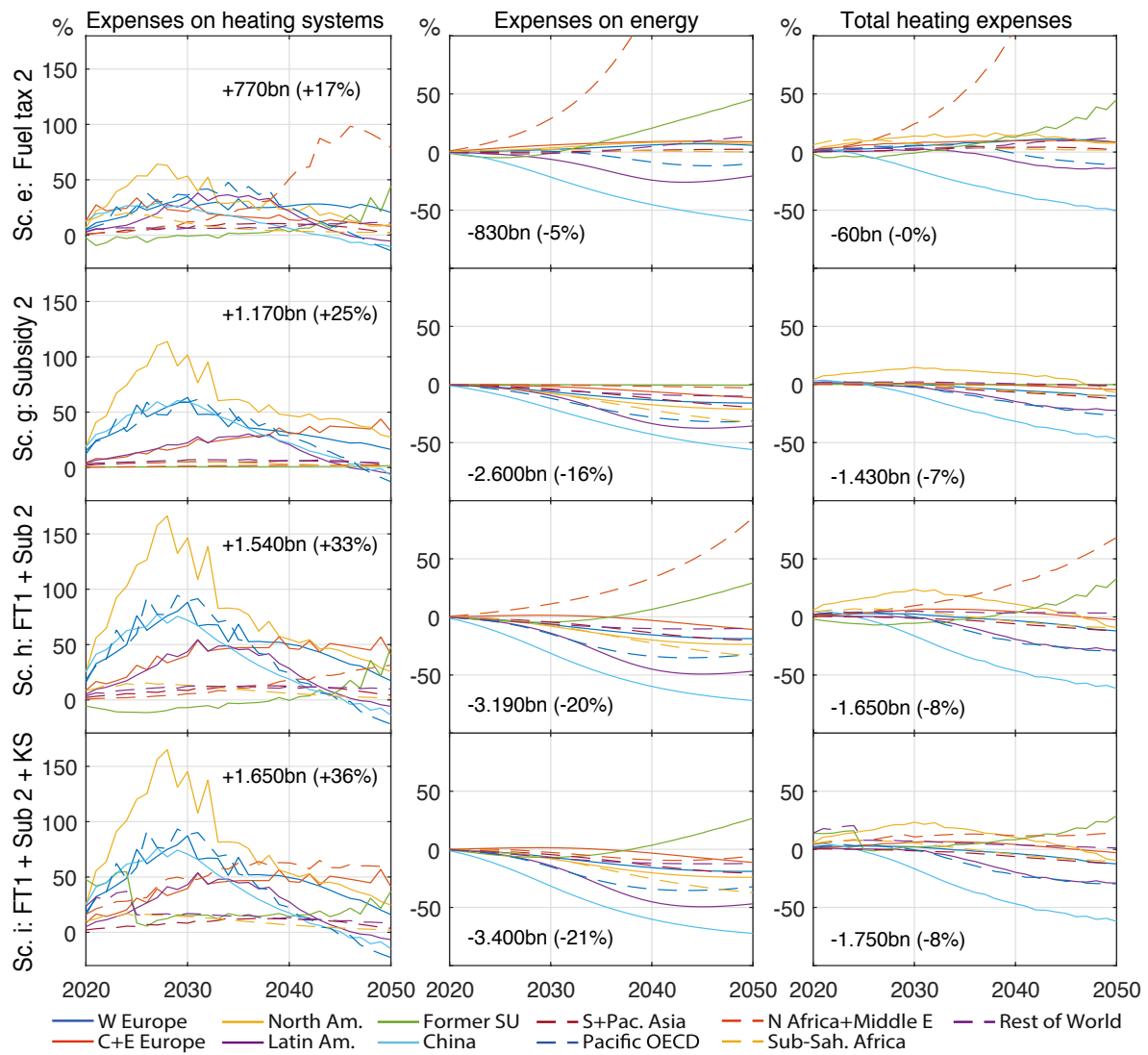

**Figure 4** Changes in expenses on residential heating (in %) per world region, for exemplary policy scenarios (aiming at uptake of low-carbon technologies), relative to scenario c (improved insulation and retrofitting, without technology policies). Monetary values show the cumulative global changes in each panel (in constant 2015-Euros), summed over 2020-2050.

In all scenarios, projected savings from energy expenses exceed the additional costs for the purchase of new heating systems (assuming constant energy prices), leading to reductions in overall expenses on heating by up to 8%. The notable exemption is scenario **j**, in case of which the transition to (relatively inefficient) electric resistance heating would increase overall expenses by more than 20%. Projected net savings are largest for policy mixes which involve both taxes and subsidies (**h** and **i**). While they show the largest increases in expenses on heating systems, those would also enable the largest energy savings and emission reductions.

Households, however, do not directly face the changes in net costs. They also need to pay for the carbon taxes, while potentially benefiting from purchase subsidies. When taxes are used as the only policy instrument, tax payments would by far exceed the achievable savings in real costs. Net benefits for households would then depend on the way in which tax revenues are redistributed. In case of policy mixes, part of the tax revenues would be recycled into purchase subsidies. Savings in real costs would then exceed the net burden from policies to households (tax payments minus subsidy payments) by a factor of 3-7.

Importantly, costs and savings would not occur simultaneously (see figure 4). In most world regions, additional expenses on heating systems would peak around 2030. Meanwhile, resulting changes in energy expenses only gradually increase over time, not reaching their full extent before 2050. In most world regions, substantial net savings are thus not projected before 2040. This type of temporal trade-off implies that while overall net savings may be large from a public policy perspective, specific technology choices may still remain unattractive from the perspective of individual households (which tend to discount future savings), if they are not incentivised by policies.

Furthermore, costs and savings are not equally distributed between world regions. While most regions show a similar profile of relative changes in expenses over time, some regions are projected to realise much larger net savings (e.g., in China), while others could face net cost increases for longer periods of time (e.g., in North America). The largest relative cost increases could occur in Russia and Nord Africa/Middle East, where current fossil fuel prices are considerably lower than in other world regions. The addition of 'kick start' policies for those regions, however, is projected to reduce those cost increases considerably, as it would incentivise the diffusion of more efficient alternative technologies.

## 4.4  Dynamics in an exemplary decarbonisation scenario

We here analyse the dynamics of decarbonising residential heating, focusing on scenario **i**. The cost dynamics underlying the technology transition are illustrated in Figure 5, and the resulting impacts on technological change in Figure 6, under different assumption regarding household behaviour.

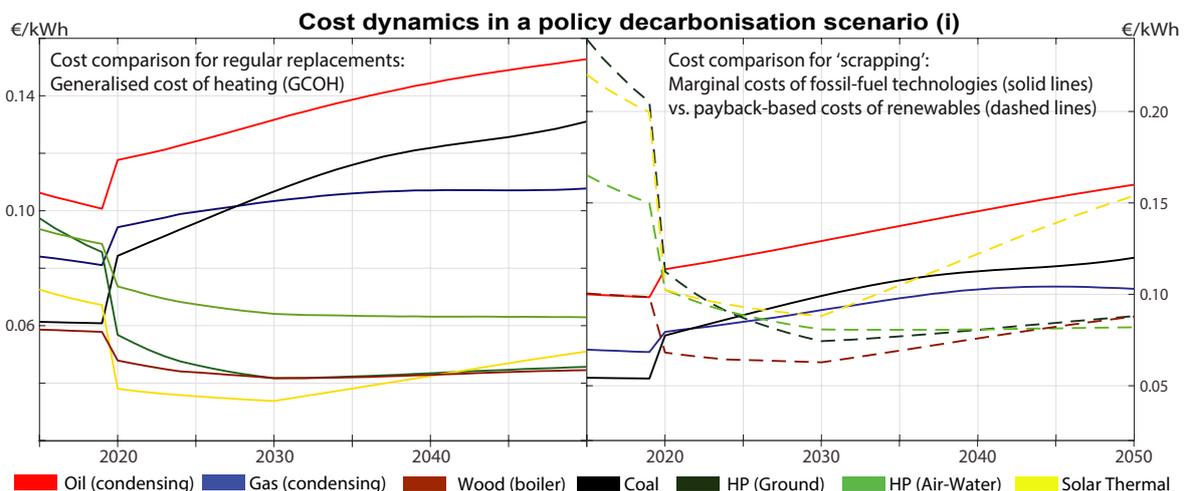

**Figure 5** Technology cost dynamics in case of decarbonisation policy scenario i. Left: generalised cost of heating per technology, incl. policies, which are the basis of regular replacement decisions by households. Right: marginal running costs of fossil fuel systems compared to payback based costs of renewables (for a three-year payback period), the comparison of which forms the basis for modelled 'scrapping' decisions.

The left panel of Figure 5 shows global averages for the generalised cost per technology as seen by households, including policies and 'intangibles'. Without new policies and at current prices, heat pumps are not yet competitive with gas heating, but on par with oil. Solar thermal is already competitive on average, which mainly reflects low costs in China, where 66% of the global capacity was installed in 2014. Biomass is competitive with fossil fuels in most world regions, but uptake is often less influenced by prices than by households' preferences, which tend to regard biomass as less convenient. When introducing new policies in 2020, the carbon tax increases the cost of fossil fuel heating, while subsidies decrease the cost of renewables, making them competitive with gas. The gradual phase-out of subsidies after 2030 is largely compensated by learning-induced decreases in investment costs, keeping levelised costs of renewables relatively stable until 2050. Induced by large capacity increases, investment costs for solar thermal and ground-source heat pumps are projected to decrease by -20% and -45% until 2030 (relative to 2014), respectively, with further reductions until 2050 (overall -30% and -66%, respectively). For comparison: the IEA (2016b) expects the costs of solar thermal to decrease by around -40% until 2030.

The right panel shows the marginal running costs of fossil fuel heating, compared to full payback costs for renewable technologies (assuming a payback period of three years). Choice experiments on the behavioural decision-making of households indicate that as long as the former do not exceed the latter, households would not consider a premature replacement, even if it would be highly profitable from an outside perspective (Newell and Siikamäki, 2015; Olsthoorn et al., 2017). The combined effect of the 50-200€/t$CO_2$ fuel and 50% subsidy on renewables is just large enough (despite translating into much larger differences in levelised costs): after 2020, the fuel savings from replacing an oil or gas system by renewables increasingly exceeds the necessary investment in less than three years, potentially incentivizing the scrapping of fossil fuel capacities before they reach the end of their rated technological lifetime.

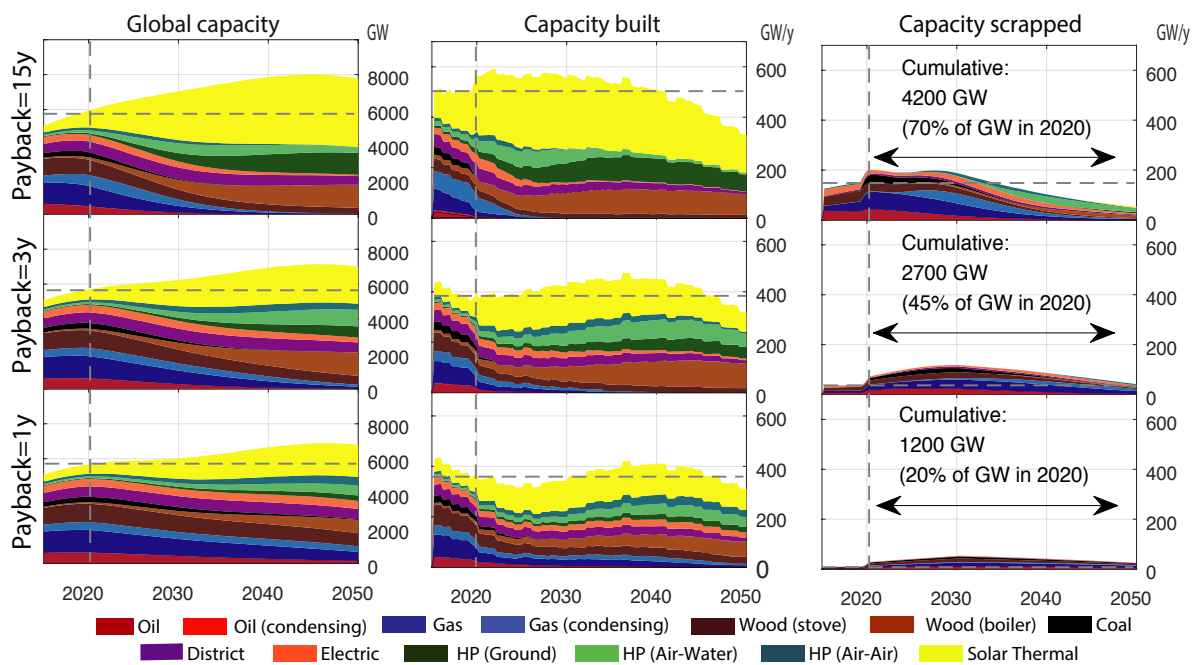

**Figure 6** Replacement dynamics of global heating capacity under different behavioural assumptions on the average payback threshold for premature replacements: 15 years (SD=5y), 3 years (SD=1y), and 1 year (SD=1/3y). Panels show the stock of global heating capacity (left), heating capacity which is built within each year (centre), and heating capacity which is scrapped prematurely within each year (right). Horizontal dashed lines represent 2014 levels, vertical dashed lines the introduction of simulated policies in 2020.

The resulting dynamics in technological change can be seen in Figure 6, which shows projections for the total global heating capacity, capacity built per year, and capacity scrapped per year, under different behavioural assumptions. Under the modelled mean payback threshold of 3 years (middle panels), scrapping happens rarely before 2020: less than 20GW (0,3%) of the global capacity would be replaced prematurely each year, while total capacity additions are around 400GW/y (6,6%). After 2020, scrapping would gradually increase, peaking at 115GW/y (around 1,8%) in the 2030s. While small on an annual basis, the model suggests that households would scrap a total capacity of 2.700GW from 2020-50, which is around 45% of the installed capacity in 2020. The induced dynamics are self-reinforcing, allowing a faster growth of households' experience and industry know-how regarding renewable heating technologies, and eventually an almost complete phase-out of fossil fuel systems by 2050.

The importance of household behaviour and related uncertainty becomes evident in comparison with the upper and lower panels of Figure 6, which depict the same dynamics for hypothetical mean payback thresholds of 15 years and 1 year. Applying a threshold of 15 years is equivalent to a discount rate of 3% (given a technical life expectancy of 20 years), implying the optimistic assumption that household decisions take into account potential future savings to almost full extent. If households were indeed acting in that way, the model suggests that they should scrap around 130GW/y (2,4%) even before policies get introduced in 2020, and up to 200GW/y (3,6%) immediately afterwards. Households would also favour more capital-intensive technologies, as evidenced by the rapid uptake of solar thermal and ground-source heat pumps. In a world of such forward-looking households, heating could be largely decarbonised by 2035, around 15 years earlier than in scenario **i**.

The opposite result is projected if households were to apply an average payback threshold of 1 year. In this case, they would only replace their heating system once expected savings exceed the upfront cost within one year. Scrapping would then be virtually non-existent before 2020 (8GW/y), and not exceed 50GW/y (0,8%) in the 2030s. While renewables could still grow to service additional demand (such as in China), the decarbonisation of the existing building stock would largely depend on regular replacements. Given average technical lifetime of 20 years and the implemented model assumptions on diffusion dynamics (in which growth is correlated to current market shares), our projections suggest that a complete transition is then not achievable until 2050, at least not under the simulated policies: fossil fuel capacities in 2050 would remain at 40% of their current level.

# 5 Discussion and Conclusion

Our results suggest that an almost-complete decarbonisation of residential heating is possible until 2050, based on a combination of improved building insulation and existing technologies, but unlikely to happen without stringent policy instruments. Policy mixes are projected to be more effective than a carbon tax on its own for driving the market of low-carbon technologies, involving lower cumulative net emissions and reduced cost burdens for households. From a societal perspective, emission reductions may be achieved at low or even negative cost: initially higher capital expenses could lead to permanently decreased energy expenses in most world regions.

The simulated technology transitions in residential heating would need decades rather than years, in parts simply due to long average lifetimes of heating equipment. Given such time-scales, our model projections suggest that a complete decarbonisation until 2050 does not only require an immediate ramp-up of low-carbon investments, but also that households replace (or 'scrap') a substantial share of inefficient heating systems prematurely. Therefore, the simulated effectiveness of policies inevitably depends on behavioural assumptions on 'scrapping' decisions.

Overall, there remains a considerable degree of uncertainty regarding behaviour, data and the future development of technology characteristics, under which the true long-term effect of any policy is hard to estimate a priori. Representing all relevant behavioural factors in a quantitative global energy model may remain an unachievable benchmark, given that no model can ever be more than a stylised representation of reality. We thus chose a midway compromise, integrating in a stylised form available knowledge on technology diffusion and household decision-making into a bottom-up simulation model of technology choice. While it remains limited in its degree of realism, we argue that it provides a clear improvement on incumbent optimisation models, shedding light on important diffusion dynamics and behavioural uncertainties.

Our results show that such uncertainties are particularly relevant in the context of limiting global warming to 1.5°C, which requires policies aiming at rapid deep decarbonisation, and are outside of what has so far been implemented in most parts of the world. Due to our inclusion of behavioural features, however, our projections are potentially more valuable to policy-makers for carrying out impact assessments of possible sets of policies, in comparison to standard optimization models that assume perfect information and rationality. Due to their use of unrealistic behavioural assumptions, the latter could potentially mislead policy-makers towards excessively simplistic policy strategies for incentivising the decarbonisation of households.

Other aspects of household decision-making are likely relevant, but still remain unspecified in our modelling - such as split incentives (e.g. in case of rented property), or a limited access to finance (which is one possible reason explaining low required payback times). The value of 'intangibles', which we estimate from historical diffusion trends, are not necessarily constant over decades, but may change over time. Some behavioural uncertainties remain impossible to simulate, so that our results may still be considered as optimistic.

While our modelling achieves the target with our set of assumed behavioural features, it can only indicate the potential of behaviourally-oriented policies. Much remains unknown on how to specifically design and implement such policies, which must take into account as much additional behavioural knowledge as possible, and can benefit substantially from psychological and sociological research. While the evidence base remains relatively thin, there is little time to spare, and therefore further research in the direction of behavioural science and modelling will need to be carried out in conjunction to the introduction of policies.

# 6 Appendix
## 6.1 Tables

**Appendix-Table 1** Grouping of the 59 E3ME regions (right) into eleven major world regions (left) (aggregation is only for presentation of results, FTT:Heat simulates all 59 regions individually).

| | |
|---|---|
| **Western Europe** | Belgium, Denmark, Germany, Greece, Spain, France, Ireland, Italy, Luxembourg, Netherlands, Austria, Portugal, Finland, Sweden, UK, Cyprus, Malta, Norway, Switzerland, Iceland, Turkey |
| **Central and Eastern Europe** | Czech Republic, Estonia, Latvia, Lithuania, Hungary, Poland, Slovenia, Slovakia, Bulgaria, Romania, Croatia, Macedonia |
| **North America** | USA, Canada |
| **Latin America** | Mexico, Brazil, Argentina, Colombia, Rest of Latin America |
| **Former Soviet Union** | Russia, Belarus, Ukraine |
| **China** | China |
| **South and Pacific Asia** | Indonesia, Rest of ASEAN, India, Korea, Taiwan |
| **Oceania (Pacific OECD)** | Australia, New Zealand, Japan |
| **North Africa and Middle East** | Saudi Arabia, OPEC excl. Venezuela, Africa OPEC |
| **Sub-Saharan Africa** | Nigeria, South Africa, Rest of Africa |
| **Rest of World** | Rest of World (all other countries) |

## 6.2 Estimation of intangible cost components

The technology- and region-specific empirical parameter $\gamma_i$ is estimated in a two-step calibration procedure (shown in Appendix-Figure 1): first, we run the model based on the cost estimates of $LCOH_i$ only, with $\gamma_i=0$ for all technologies. For each region, we then compare the growth of technologies as projected by the model with the diffusion trend as observable in the historic data, using a graphical interface. If there are deviations, we iteratively adjust the values of $\gamma_i$ (upwards or downwards), until the projected diffusion at the start of the simulation is consistent with the empirical trends. The estimated values of 'intangibles' are assumed to be constant over the simulation period, since we have no possibility to estimate future preferences.

On average, we find that oil, gas (incl. LPG) and district heating are slightly more attractive to households than suggested by the pure costs, leading to values of $\gamma_i$ in the range of -10% to -15%, relative to the pure $LCOH_i$. For electric resistance heating, the average $\gamma_i$ is equivalent to -30% of its $LCOH_i$. The opposite holds for coal and biomass heating, for which typical $\gamma_i$ values are between +40% and +80%. For solar thermal and heat pumps, values differ by region, usually between -20% and +20%.

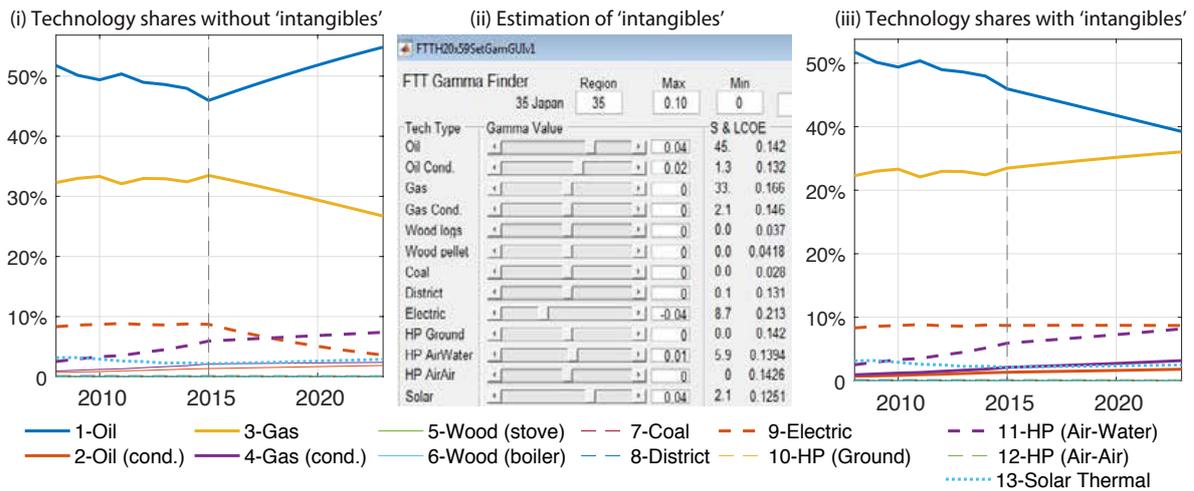

**Appendix-Figure 1** Estimation of 'intangible' cost components with FTT:Heat, at the example of Japan. (i) shows the historic (2008-2014) and projected trends in technology shares (2015-2022) without 'intangibles', (ii) shows the graphical interface for adjusting the values of 'intangibles', (iii) shows the resulting trends in technology shares with the 'intangibles'. The dashed vertical lines indicate the start of the model simulation.

## 6.3  Detailed technology data

**Appendix-Table 2** Model assumptions for residential heating technologies. Costs refer to mean values. (Data sources: Fleiter et al. (2016), IEA/ETSAP (2012), Danish Energy Agency (2013), EHPA (2016)).

|  | Upfront cost (€/kW$_{th}$) | O&M cost (€/kW$_{th}$ pa) | Efficiency (kWh$_{th}$/kWh) | Learning rate (%) |
| --- | --- | --- | --- | --- |
| Oil | 471 | 19 | 0.75 | --- |
| Oil condensing | 512 | 20 | 0.86 | -10% |
| Gas | 391 | 8 | 0.75 | --- |
| Gas condensing | 434 | 9 | 0.9 | -10% |
| Biomass stove | 440 | 0.1 | 0.1-0.7 | --- |
| Biomass boiler | 523 | 2 | 0.85 | -10% |
| Coal | 247 | 5 | 0.75 | --- |
| District heating | 265 | 16 | 0.98 | --- |
| Direct electric | 538 | 0.5 | 1.00 | --- |
| HP- ground source | 1400 | 14 | 3.50 | -30% |
| HP- air/water | 750 | 15 | 2.50-2.70 | -30% |
| HP- air/air | 510 | 51 | 2.50-2.70 | -30% |
| Solar thermal | 773 | 8 | --- | -10% |

Conversion efficiencies refer to the ratio of thermal energy 'leaving' the heating system, relative to the necessary energy input, covering both space and water heating. In case of traditional biomass, lower conversion efficiencies (10-50%) are assumed in developing countries (IEA, 2014). For heat pumps, efficiency values are defined as their seasonal performance factor (the annual average ratio of delivered heat to electricity input), which differs by climate region. For solar thermal, local productivities are calculated from data by the IEA Solar Heat Programme (Mauthner et al., 2016), which we integrate into the model as region-specific capacity factors (i.e., units of heat produced per unit of capacity installed). The average technical life expectancy is set to 20 years for all technologies, based on literature estimates (the actual lifetime may be shorter if households decide to replace a system earlier for economic reasons) (Danish Energy Agency, 2016; IEA/ETSAP, 2012).

**Appendix-Table 3** Model estimates for levelised cost of heating (LCOH), generalised cost of heating (GCOH), estimated values of intangible cost components, and standard deviation of LCOH. All numbers refer to mean values, calculated as the average of all 59 modelled world regions (weighted by their heat demand).

|  | LCOH | GCOH | Estimated values of intangibles | | Standard deviation of LCOH | |
| --- | --- | --- | --- | --- | --- | --- |
|  | €-Cent/kWh$_{th}$ | €-Cent/kWh$_{th}$ | €-Cent/kWh$_{th}$ | %LCOH | €-Cent/kWh$_{th}$ | %LCOH |
| Oil | 11,1 | 9,7 | -1,3 | -12% | 0,02 | 17% |
| Oil condensing | 10,2 | 9,0 | -1,3 | -13% | 0,02 | 17% |
| Gas | 9,3 | 8,0 | -1,2 | -13% | 0,02 | 17% |
| Gas condensing | 8,2 | 6,6 | -1,6 | -20% | 0,01 | 18% |
| Biomass stove | 5,8 | 5,1 | -0,7 | -11% | 0,02 | 29% |
| Biomass boiler | 4,3 | 5,9 | 1,6 | 37% | 0,01 | 30% |
| Coal | 3,0 | 5,5 | 2,5 | 84% | 0,01 | 27% |
| District heating | 7,2 | 6,2 | -1,1 | -15% | 0,01 | 16% |
| Direct electric | 11,8 | 8,6 | -3,2 | -27% | 0,02 | 14% |
| HP- ground | 9,3 | 8,7 | -0,7 | -7% | 0,02 | 25% |
| HP- air/water | 8,7 | 8,0 | -0,7 | -8% | 0,02 | 21% |
| HP- air/air | 8,8 | 8,4 | -0,4 | -5% | 0,02 | 19% |
| Solar thermal | 10,8 | 10,0 | -0,8 | -8% | 0,04 | 33% |

## 6.4 Regional technology pathways in scenario i

Appendix-figure 2 summarises the simulated future development of residential heating systems in five major world regions, together accounting for 80% of heat demand and direct heating $CO_2$ emissions in 2014. Technology composition and emissions by region are shown for a projection under improved levels of building insulation (scenario **c**) on the left, and under additional policies for technological decarbonisation of heating (scenario **i**) on the right.

In Western Europe and North America, heating is currently dominated by gas. Under baseline conditions, their technology composition would stay relatively constant, while the ongoing diffusion of more efficient gas heating systems and a limited uptake of renewables implies a continued, but slow decrease of emissions. Under policies in scenario h, fossil fuel heating and direct electric heating in both regions would be substituted by a mix of heat pumps and solar thermal, with only limited diffusion of modern biomass.

In Central and Eastern Europe, large shares of the heating demand are currently serviced by district heat networks, biomass and gas. Most direct emissions, however, originate in a relatively large share of coal heating, compared to other world regions. Modern renewables are mostly absent under current technology diffusion trends. Under decarbonisation policies, our model suggests that fossil fuels would be substituted by a mix of biomass and heat pumps, while the projected uptake of solar thermal remains low.

In China, fossil fuels still play a smaller role in heating. Historically, most emissions result from coal. Oil and gas are on the rise, however, and would continue their ongoing growth under baseline conditions. At the same time, a widespread uptake of solar thermal systems and heat pumps is taking place, so that baseline emissions are projected to peak around 2030 and decrease afterwards. With additional policies, the shift to renewables would be accelerated, with solar supplying the equivalent of two-thirds of the water heating demand by 2050. Importantly, China is also the region with largest projected demand increases until 2050, driven by continuously rising incomes (and coming from historically very low values of heating intensity).

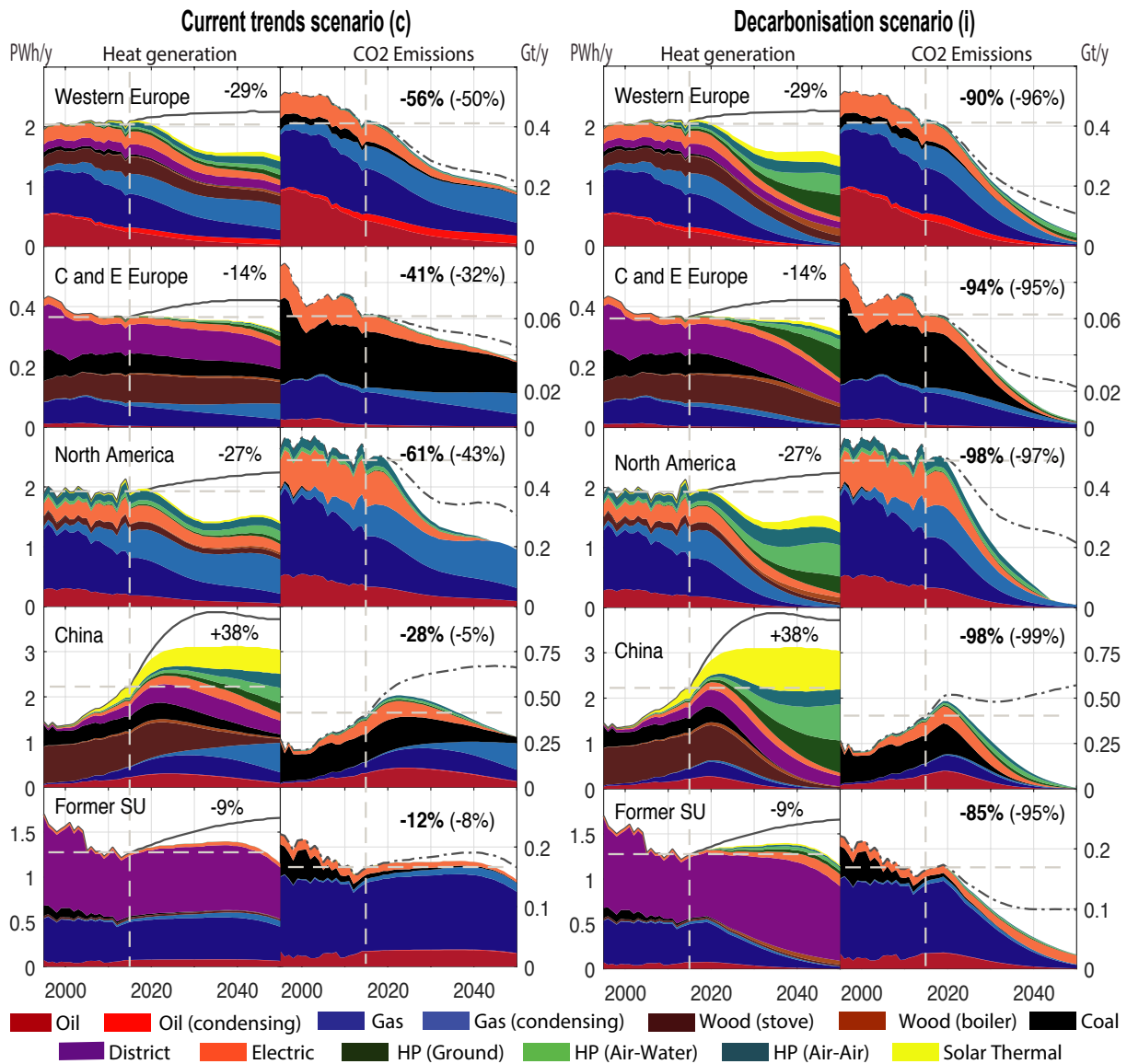

**Appendix-Figure 2** Technology composition and $CO_2$ emissions (direct on-site and indirect electricity emissions) in the residential heating sector in five world regions, under current technology trends with improved insulation and retrofitting (c) and a scenario with -95% global direct decarbonisation by 2050 (i). Model simulations by FTT:Heat start in 2015 (indicated by vertical dashed lines). Horizontal dashed lines represent 2014 levels. Solid curves show the baseline demand trends from scenario a, and dashed curves the total emissions should the power sector not be decarbonised. Percentage values refer to the change by 2050, relative to 2014. Bold percentage values indicate the reduction in annual total $CO_2$ emissions (direct + indirect), the values in brackets show the corresponding reduction in direct on-site $CO_2$ emissions.

In the Former Soviet Union, 60% of heat is currently provided by district heating systems, the remaining fraction by gas and oil. No use of renewables is reported, limiting the potential for a fast technology transition. Therefore, only little change would take place in the baseline, with slowly rising emission levels until 2050. With decarbonisation policies, only the introduction of 'kick start' schemes would eventually lead to a limited uptake of renewables. Most emissions occur not on site, but in centralized heat plants. A parallel decarbonisation would thus need to take place in the country's district heating systems, which are not explicitly modelled here (the same holds for central heating in other world regions).